\documentclass[manuscript=article,maxauthors=10]{achemso}
\setkeys{acs}{etalmode=truncate,maxauthors=10}

\usepackage{chemformula}
\usepackage[T1]{fontenc}
\usepackage{undertilde}
\usepackage{hyperref}
\usepackage[symbol]{footmisc}

\newcommand{\kms}{km s$^{-1}$}

\newcommand{\h}{^{\mbox{\scriptsize h}}}
\newcommand{\m}{^{\mbox{\scriptsize m}}}

\usepackage{xcolor}
\usepackage{multirow}
\usepackage{amssymb}

\usepackage{comment}
\usepackage{subcaption}

\newcommand{\arcdeg}{^\circ}
\newcommand{\arcmin}{^\prime}
\newcommand{\arcsec}{^{\prime\prime}}
\newcommand{\farcs}{.\!\!^{\prime\prime}}
\newcommand{\fs}{.\!\!^{\mbox{\scriptsize s}}}

\author{Olivia H. Wilkins}
\affiliation{Division of Chemistry and Chemical Engineering, California Institute of Technology, Pasadena, CA 91125 USA}
\altaffiliation{Current address: NASA Postdoctoral Program Fellow, NASA Goddard Space Flight Center, Greenbelt, MD, 20771 USA}
\email{olivia.h.wilkins@outlook.com}

\author{Geoffrey A. Blake}
\affiliation{Division of Chemistry and Chemical Engineering, California Institute of Technology, Pasadena, CA 91125 USA}
\alsoaffiliation{Division of Geological and Planetary Sciences, California Institute of Technology, Pasadena, CA 91125 USA}

\title{On the Relationship between \ch{CH3OD} Abundance and Temperature in the Orion KL Nebula}

\begin{document}

\begin{abstract}

The relative abundances of singly-deuterated methanol isotopologues, \textcolor{black}{[\ch{CH2DOH}]/[\ch{CH3OD}]}, in star-forming regions deviate from the statistically expected ratio of 3. In Orion KL, the nearest high-mass star-forming region to Earth, the \textcolor{black}{singly-deuterated} methanol ratio is about 1, and the cause for this observation has been explored through theory for nearly three decades. We present high-angular resolution observations of Orion KL using the Atacama Large Millimeter/submillimeter Array to map small-scale changes in \ch{CH3OD} column density across the nebula, which provide a new avenue to examine the \textcolor{black}{deuterium} chemistry \textcolor{black}{during} star and planet formation. By considering how \ch{CH3OD} column densities vary with temperature, we find evidence of chemical processes that can significantly alter the observed column densities. The astronomical data are compared with existing theoretical work and support \textcolor{black}{D-H exchange} between \ch{CH3OH} and heavy water \textcolor{black}{(i.e., HDO and \ch{D2O})} at methanol's hydroxyl site in the icy mantles of dust grains. 
\textcolor{black}{The enhanced \ch{CH3OD} column densities are localized to the Hot Core-SW region, a pattern that may be linked to the coupled evolution of ice mantel chemistry and star formation in giant molecular clouds.} This work provides new perspectives on deuterated methanol chemistry in Orion KL and informs considerations that may guide future theoretical, experimental, and observational work.

\end{abstract}

\section{Introduction} \label{sec:introch3od} 

\textcolor{black}{The relative abundances of site-specific stable isotopologues}, particularly those involving deuterated compounds, are powerful tools that can be used to trace chemical evolution in the interstellar medium, and \textcolor{black}{during} star and planet formation. For example, the relative abundances of heavy water \textcolor{black}{(i.e., HDO and \ch{D2O})} in comets and meteorites can provide insights into the processing of water between the primordial molecular cloud and present-day Earth.\citep{Altwegg2015,Altwegg2017} \textcolor{black}{Interstellar compounds such as \ch{N2H+} and \ch{CH3OH}} have D/H ratios that are higher than the cosmic value of ${\sim}10^{-5}$,\citep{Zavarygin2018} and that ratio is a function of temperature, with higher D/H ratios signalling formation in colder, typically denser, environments.\citep{Fontani2015}

In many high-mass and low-mass protostars alike, the deuterium chemistry of methanol---namely, the relative abundances of the singly-deuterated isotopomers \ch{CH2DOH} and \ch{CH3OD}---has presented itself as a mystery. Methanol is one of the simplest complex (having $\ge$6 atoms) organic molecules, and it is found at every stage of star formation, from cold cloud cores and hot cores/corinos to outflows and circumstellar disks. \citep{Herbst2009,Walsh2016} As such, it is commonly used as a tracer of other complex organics. 

In prestellar and protostellar cores, methanol forms primarily via successive hydrogenation of frozen CO on grain mantles (eq~\ref{eq:ch3oh}).\citep{Woon2002,Watanabe2002}
\begin{equation}\label{eq:ch3oh}
    \ch{CO ->[H] HCO ->[H] H2CO ->[H] H3CO ->[H] CH3OH}
\end{equation}
Statistically, we would expect the [\ch{CH2DOH}]/[\ch{CH3OD}] ratio to be 3 since there are three methyl hydrogen sites compared to a single hydroxyl site. This statistical ratio has been observed in the massive star-forming region NGC 7538-IRS1 but not toward many other star-forming regions.\citep[][]{OspinaZamudio2019} Low-mass cores generally exhibit ratios ${>}3$ and as much as $\utilde{>}$10,\citep{Bizzocchi2014,Taquet2019} while high-mass \textcolor{black}{protostars} tend to have [\ch{CH2DOH}]/[\ch{CH3OD}] ratios of ${<}3$.\citep{Ratajczak2011,Belloche2016, Bogelund2018} \textcolor{black}{Astrochemical models have predicted that the [\ch{CH2DOH}]/[\ch{CH3OD}] ratio should be $\ge$10 in prestellar cores and that \ch{CH3OD} is only efficiently formed on icy grains at later evolutionary stages when the ices are warmed due to the presence of young (proto)stars.\citep[][]{Kulterer2022}} Deviations from the statistical ratio \textcolor{black}{in high-mass star-forming regions} have \textcolor{black}{also} been attributed to grain surface chemistry,\citep{Parise2005, Ratajczak2009} but investigations into the intricacies of such processes---and the potential role of gas processing---are ongoing.

The Orion Kleinmann-Low (Orion KL) nebula is a high-mass star-forming region notable for its peculiar methanol deuteration. \textcolor{black}{At a distance of $\sim$388 pc,\citep[][]{Kounkel2017} Orion KL is uniquely situated to explore the relationship between relative deuterated methanol abundances and environmental conditions because sub-environments within the nebula can be resolved, even with modest imaging capabilities. The two most well-studied regions within Orion KL are the Hot Core and Compact Ridge. The Hot Core region contains denser and warmer gas ($n_{\rm H_2}\sim 10^7$ cm$^{-3}$, $T_{\rm kin}\sim 200$ K) whereas the Compact Ridge, to the southwest,\footnote{In astronomical maps, north is up, east is left, and west is right.} is cooler and less dense ($n_{\rm H_2}\utilde{<}10^6$ cm$^{-3}$, $T_{\rm kin}\sim 100{-}150$ K).\citep[][]{Blake1987,Genzel1989} These regions are also the prominent sites of nitrogen-bearing and oxygen-bearing compounds, respectively.\citep[][]{Blake1987,Friedel2011} Extending southwest from the Hot Core toward the Compact Ridge is the Hot Core-SW, which is physically and chemically heterogeneous, with possible sources of internal heating\citep[][]{Wilkins2022} and both oxygen- and nitrogen-bearing compounds.\citep[][]{Friedel2008,Favre2011} Flanking these regions are compact sources, such as Source I---an edge-on disk thought to be internally heated\citep[][]{Wright2020,Wright2022}---to the west within the Hot Core region and the (sub)millimeter sources SMA1 and C22---a protostar and possible hot core, respectively\citep[][]{Beuther2006,Friedel2011}---to the southwest of Source I and along the northwestern edge of the Hot Core-SW. }

\citet{Jacq1993} reported the first definitive detection of \ch{CH2DOH} toward Orion KL \textcolor{black}{on angular scales between 12$^{\prime\prime}$ and 26$^{\prime\prime}$ (centered on the Hot Core and Source I region). They} combined their measurements with past \ch{CH3OD} measurements to report a [\ch{CH2DOH}]/[\ch{CH3OD}] ratio in the range of 1.1-1.5. \citet{Neill2013} similarly reported a ratio of $1.2\pm0.3$ \textcolor{black}{based on local thermodynamic equilibrium models of ${\sim}30{-}44^{\prime\prime}$ observations of the Hot Core and the Compact Ridge}, while  
\citet{Peng2012} reported an even lower ratio of $0.7\pm0.3$ toward \textcolor{black}{Orion KL, using an angular resolution of $3\farcs6\times2\farcs3$}.

There has been extensive debate about whether the apparent \ch{CH3OD} enhancements are the result of grain-surface or gas-phase processes. Early on, \citet{Jacq1993} concluded that their observed ratio was evidence of grain-surface processing followed by injection into the gas phase\textcolor{black}{, perhaps by thermal desorption}. Shortly after, chemical models of gas-phase exchange rejected the grain-surface hypothesis on the premise that such chemistry would require unrealistically high [HDO]/[\ch{H2O}] ratios.\citep[][]{Charnley1997} 
\citet{Rodgers2002} criticized the assumed statistical [\ch{CH2DOH}]/[\ch{CH3OD}] ratio of 3 since D and H react with species other than CO and \ch{H2CO}, which could affect the relative abundances of the singly-deuterated methanol isotopologues. \citet{Osamura2004} used models to suggest that ion-molecule reactions in the gas phase lead to the loss of \ch{CH3OD}, which they conclude accounts for high [\ch{CH2DOH}]/[\ch{CH3OD}] ratios in low-mass star-forming regions, \textcolor{black}{provided methanol is efficiently regenerated in the dissociative recombination of protonated methanol with electrons}. \textcolor{black}{This work also finds} that \textcolor{black}{D-H exchange} on the methyl site is inefficient. 

Nevertheless, \textcolor{black}{D-H exchange} \textcolor{black}{at} the hydroxyl group of methanol on icy grain mantles has emerged as a favored explanation for the [\ch{CH2DOH}]/[\ch{CH3OD}] ratios observed in massive star-forming regions. \citep{Nagaoka2005,Ratajczak2009,Peng2012} In this mechanism, deuterated water in the ice reacts with \ch{CH3OH} to produce \ch{CH3OD}:
\begin{equation}\label{eq:exchangeHDO}
    \ch{CH3OH + HDO <=> CH3OD + H2O}
\end{equation}
\begin{equation}\label{eq:exchangeD2O}
    \ch{CH3OH + D2O <=> CH3OD + HDO}
\end{equation}
However, the intricacies of this exchange are still being investigated.

This work provides a new observational perspective on the possibility of \textcolor{black}{D-H exchange} at the methanol hydroxyl site by mapping gas-phase \ch{CH3OD} abundances in Orion KL at sub-arcsecond ($\sim$0$\farcs$7) angular resolution using the Atacama Large Millimeter/submillimeter Array (ALMA)\textcolor{black}{, which corresponds to linear scales of $\sim$270 au at the nebula's distance. }
This allows us to plot gas-phase \ch{CH3OD} as a function of the local line-of-site temperature across relatively small scales within the nebula and explore temperature-dependent chemical processes that may affect the observed \ch{CH3OD} chemistry.

\section{Methods}\label{sec:methods}
\subsection{Observations}\label{sec:obsch3od} 

Observations of Orion KL were taken in ALMA Band 4 during Cycle 5 (project code: ADS/JAO.ALMA\#2017.1.01149, PI: Wilkins) on 2017 December 14, completely on the main 12-m array. The pointing center was set to  $\alpha_{\mbox{\scriptsize J2000}} = 05\h 35\m 14\fs50$, $\delta_{\mbox{\scriptsize J2000}} = -05\arcdeg 22\arcmin 30\farcs9$. These observations employed 49 antennas during one execution block. All spectra were obtained in a single local oscillator set-up consisting of 10 spectral windows; as such, the uncertainties for quantities derived from these spectra are dominated by thermal noise and mostly unaffected by calibration uncertainty. Of these spectral windows, the targeted \ch{CH3OD} lines were contained in three spectral windows with a spectral resolution of 244 kHz ($\sim$0.5 \kms) covering 143.51-143.97, 153.16-153.40, and 154.84-155.07 GHz. Projected baselines were between 15.1 m and 3.3 km (7.6 and 1650 k$\lambda$\textcolor{black}{, where $\lambda\sim2$ mm is the wavelength}), and the primary beam was 39.1$\arcsec$. The on-source integration time was 2062 s. Precipitable water vapor was 3.7 mm, and typical system temperatures were around 75-125 K. 

Calibration was completed using standard \texttt{CASA} (version 5.1.1-5) calibration pipeline scripts. {The source J0423$-$0120 was used as a calibrator for amplitude, atmosphere, bandpass, pointing, and WVR (Water Vapor Radiometer) variations, and J0541$-$0211 was used as the phase and WVR calibrator.}

The \ch{CH3OD} data introduced here were prepared in the same way as the \ch{^{13}CH3OH} images presented by \citet{Wilkins2022} In brief, the data cubes were created from measurement sets split to include only baselines of $\le$500 m, resulting in a synthesized beam of $0\farcs74\times0\farcs63$. 
Cubes were reduced with continuum emission estimated from line-free channels 
subtracted using the \texttt{uvcontsub} function followed by imaging using the \texttt{tclean} algorithm with robust weighting, a Briggs parameter of 1.5 (i.e., semi-natural weighting) for deconvolution, and the `auto-multithresh' masking algorithm \citep[][]{Kepley2020} in conjunction with interactive \texttt{tclean}. The images have a noise-level of $\sigma_{\mbox{\scriptsize RMS}}\sim 1.3$ mJy beam$^{-1}$.

\subsection{Deriving \ch{CH3OD} Parameters}\label{sec:abundance}

The \ch{CH3OD} column density as a function of position was derived via pixel-by-pixel fits of \textcolor{black}{the \ch{CH3OD} transitions shown listed in Table~\ref{tab:linesch3od}} assuming optically thin lines (see Table S1 of the Supporting Information) in local thermodynamic equilibrium (LTE).\footnote{Python script available at \href{https://github.com/oliviaharperwilkins/LTE-fit}{https://github.com/oliviaharperwilkins/LTE-fit}.} \textcolor{black}{Integrated intensity maps of each transition are providing in the Supporting Information (Figure S2).} \textcolor{black}{For each coordinate-space pixel in the data cubes, a spectrum within a single synthesized beam centered on that pixel was extracted.} The \ch{CH3OD} rotational temperature ($T_{\rm rot}$) profile was assumed to be the same as that previously derived\citep[][]{Wilkins2022} from \ch{^{13}CH3OH} since the transitions of both isotopologues have similar upper energy states $E_u$. Column density, line width \textcolor{black}{(${\sim}0.8{-}3.5$ \kms)}, and local standard of rest (LSR) velocity \textcolor{black}{(${\sim}7{-}9$ \kms)} were determined  by simultaneous fits using \texttt{LMFIT}, a least-squares fitting software package.\footnote{\href{https://doi.org/10.5281/zenodo.11813}{https://doi.org/10.5281/zenodo.11813}} Because all lines used in the fit were observed simultaneously, the uncertainties in excitation, which are derived from relative fluxes, should be dominated by thermal noise rather than by multiple sources of calibration uncertainty. 

\begin{table}
\begin{center}
\caption[Transitions of \ch{CH3OD} $v_t=0$ observed in Orion KL]{Transitions of \ch{CH3OD} used for line fits.$^\star$}\label{tab:linesch3od}
\begin{tabular}{lcccc}
\hline\hline
\multirow{2}{*}{Transition} & $\nu$ & {$E_u$} & {$S_{ij}\mu^2$} & {\multirow{2}{*}{$g_u$}}\\ 
 & {(GHz)} & {(K)} & {(Debye$^2$)} &   \\\hline
$5_{(1,4)}-5_{(0,5)}$ A & 143.7417 & 39.48 & 11.2 & 11 \\
$7_{(1,6)}-7_{(0,7)}$ A & 153.3240 & 68.05 & 14.7 & 15 \\
$3_{(-1,2)}-2_{(0,1)}$ E & 154.9628 & 17.71 & 2.2  & 7\\\hline
\end{tabular}
\end{center}
\small{\indent $^\star${From \citet{Anderson1988} Column (2): rest frequency $\nu$ of the transitions; (3): upper state energy $E_u$; (4): product of the transition line strength and the square of the electric dipole moment $S_{ij}\mu^2$; (5): upper-state degeneracy $g_u$}}
\end{table}

\section{Results and Discussion}\label{sec:resultsanddisc}

\subsection{\ch{CH3OD} Column Density}

\begin{figure}
    \centering
    \includegraphics[width=0.7\textwidth]{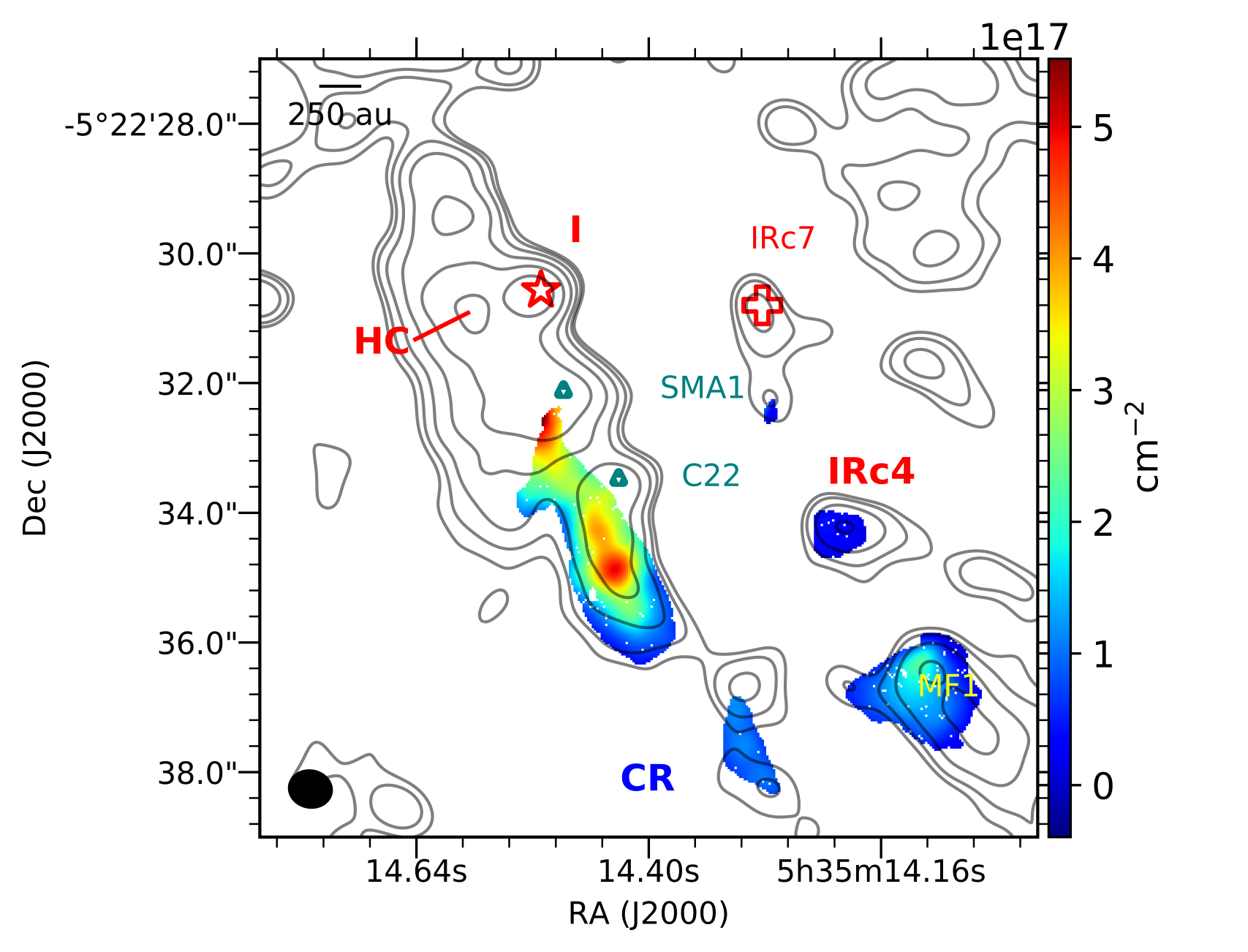}
    \includegraphics[width=0.7\textwidth]{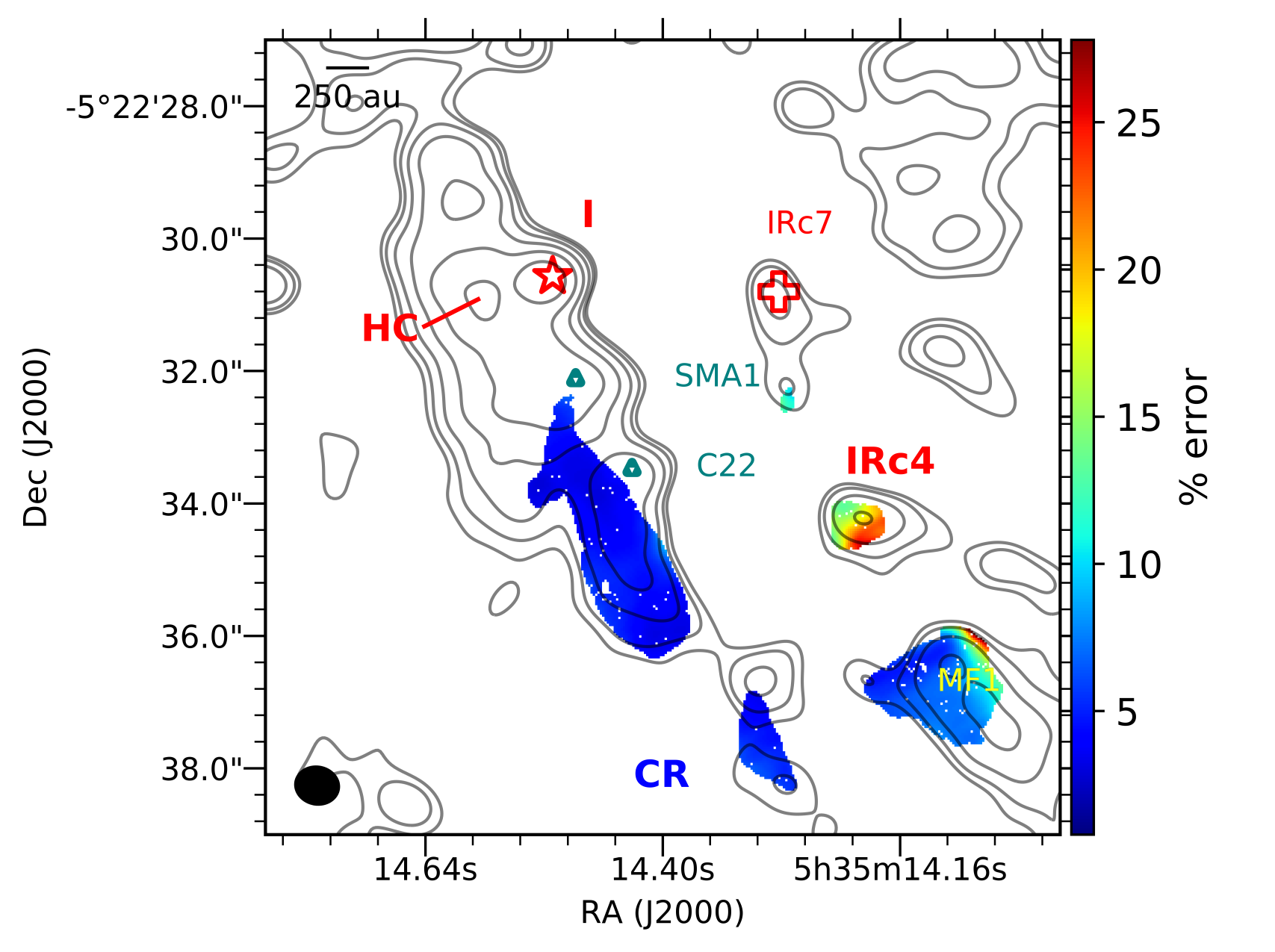}
    \caption[\ch{CH3OD} column density map]{Derived \ch{CH3OD} column density and percent propagated uncertainty shown by the color maps in the upper and lower panels, respectively. The \textcolor{black}{2 mm ($\sim$150 GHz)} continuum \textcolor{black}{emission} is shown by the grey contours at $2\sigma_{RMS}, 4\sigma_{RMS}, 8\sigma_{RMS}, 16\sigma_{RMS}, 32\sigma_{RMS}, 64\sigma_{RMS}$. \textcolor{black}{The Hot Core (HC), Source I (I), IRc7, and IRc4 are shown in red; SMA1 and C22 are shown by the teal diamonds; the methyl formate emission peak (MF1)\citep[][]{Favre2011} is labeled in yellow, and the Compact Ridge (CR) is labeled in blue. The 0$\farcs$7 synthesized beam is shown by the black ellipse in the bottom left corner.}}
    \label{fig:columnD}
\end{figure}

\begin{figure}[t]
    \centering
    \begin{subfigure}{0.495\textwidth}\centering
    \includegraphics[width=\textwidth]{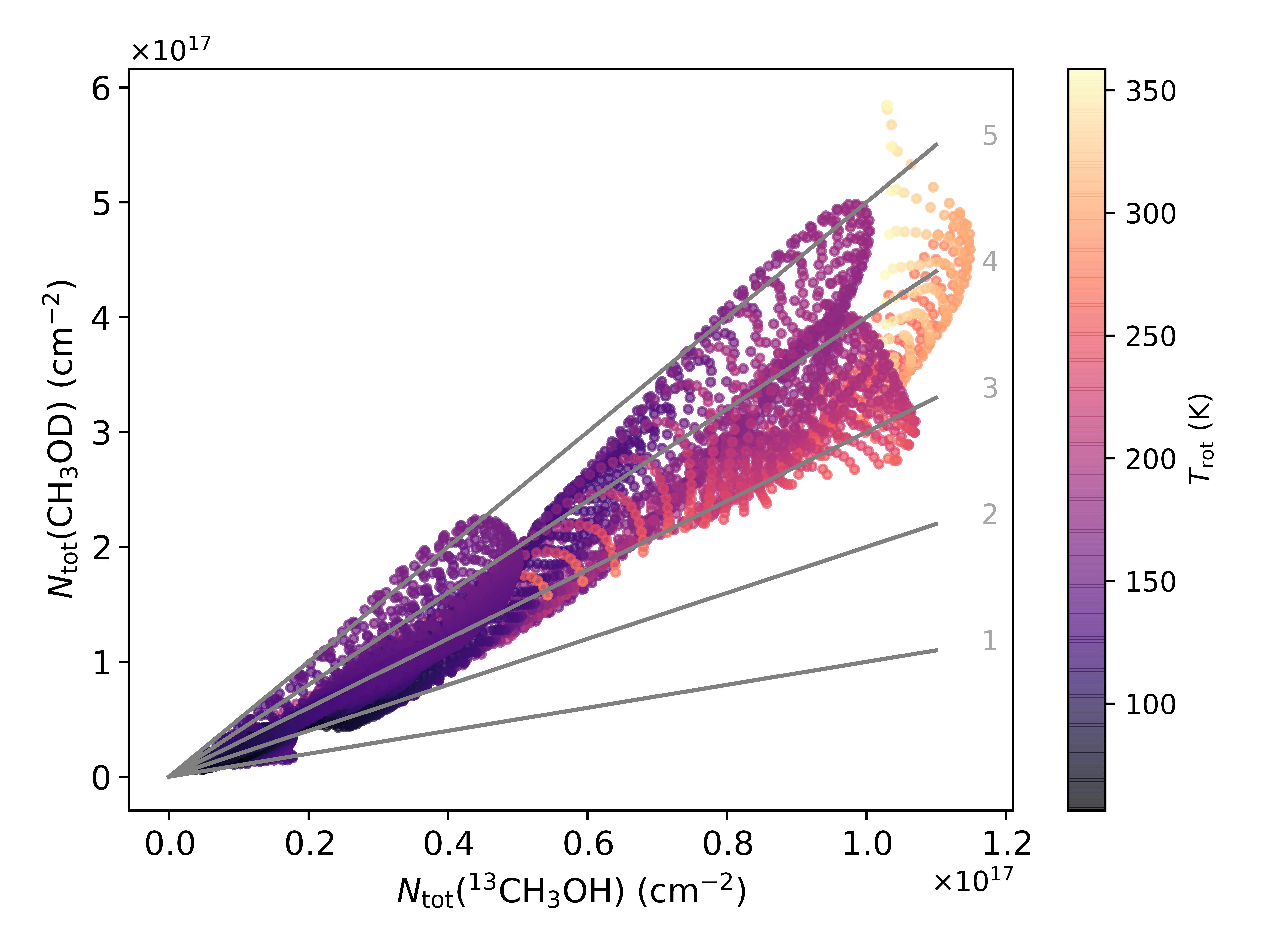}
    \caption{}    
    \end{subfigure}
    \begin{subfigure}{0.495\textwidth}\centering
    \includegraphics[width=\textwidth]{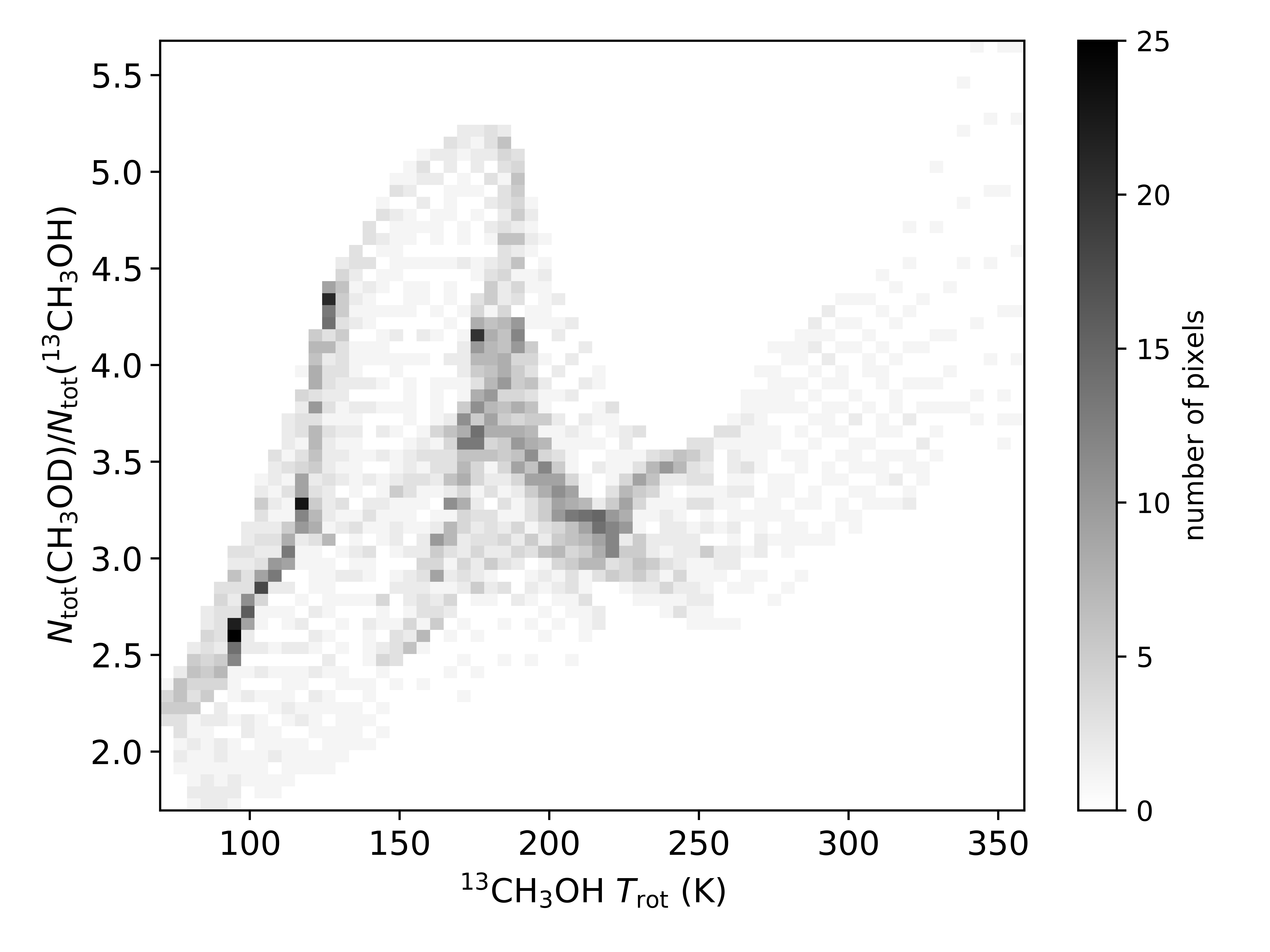}
    \caption{}
    \end{subfigure}
    \caption{\textcolor{black}{(a)} Column densities $N_{\rm tot}$ of \ch{^{13}CH3OH} (horizontal axis, from \citet{Wilkins2022}) and \ch{CH3OD} (vertical axis, this work) with each point representing a single pixel in Figure~\ref{fig:columnD}. Each pixel is colored by its rotational temperature. The grey lines are labeled by the \textcolor{black}{[\ch{CH3OD}]/[\ch{^{13}CH3OH}]} ratios (1 to 5) they represent. \textcolor{black}{(b) Two-dimensional histogram (50 points per bin) showing the methanol \textcolor{black}{[\ch{CH3OD}]/[\ch{^{13}CH3OH}]} ratios plotted as a function of temperature.}}
    \label{fig:DvsHbyTemp}
\end{figure}

The column density profile derived from a pixel-by-pixel fit of the ALMA data image cubes was used to show small-scale variations in \ch{CH3OD} column densities across Orion KL. As shown in Figure~\ref{fig:columnD}, the derived \ch{CH3OD} column density ($N_{\rm tot}$) is generally on the order of $10^{17}$ cm$^{-2}$ and peaks south of SMA1 and C22. \textcolor{black}{In general, the uncertainties (standard errors calculated using \texttt{LMFIT}) for these values are $<$10\% throughout the region southwest of the Hot Core (Hot Core-SW), which is the region of interest for most of the discussion in this work; higher uncertainties, up to $\sim$25\%, characterize IRc4 and the western (left) edge of the Compact Ridge.} The relationship between the \ch{CH3OD} and \ch{^{13}CH3OH} column densities (Figure S3 \textcolor{black}{in the Supporting Information}) and rotational temperature ($T_{\rm rot}$, \textcolor{black}{Figure S4 in the Supporting Information}) is illustrated by Figures~\ref{fig:DvsHbyTemp}a { and~\ref{fig:DvsHbyTemp}b}. In general, [\ch{CH3OD}]/[\ch{^{13}CH3OH}] falls between \textcolor{black}{1.5 and 5.3}. \textcolor{black}{Assuming a local interstellar \ch{^{12}C}/\ch{^{13}C} ratio of $68\pm15$,\citep[][]{Milam2005} this suggests [\ch{CH3OD}]/[\ch{CH3OH}] $\approx 0.015$-$0.104$, which encompasses the ratio of 0.01-0.06 in Orion KL reported by \citet{Mauersberger1988} \textcolor{black}{on much larger angular scales of 15$^{\prime\prime}$-23$^{\prime\prime}$}. The discrepancy on the higher end of these ranges may be the result of \ch{CH3OD} abundance enhancements on smaller spatial scales being diluted in the single-dish data\textcolor{black}{, but otherwise, the agreement suggests that the flux recovered in our observations is representative of that collected by single-dish observations}.  A comparison of [\ch{CH3OD}]/[\ch{^{12}CH3OH}] ratios derived from different assumed \ch{^{12}C}/\ch{^{13}C} values is presented in Table S2 of the Supporting Information.}

\begin{figure}
    \centering
    \begin{subfigure}{\textwidth}\centering
    \includegraphics[width = 0.6\textwidth]{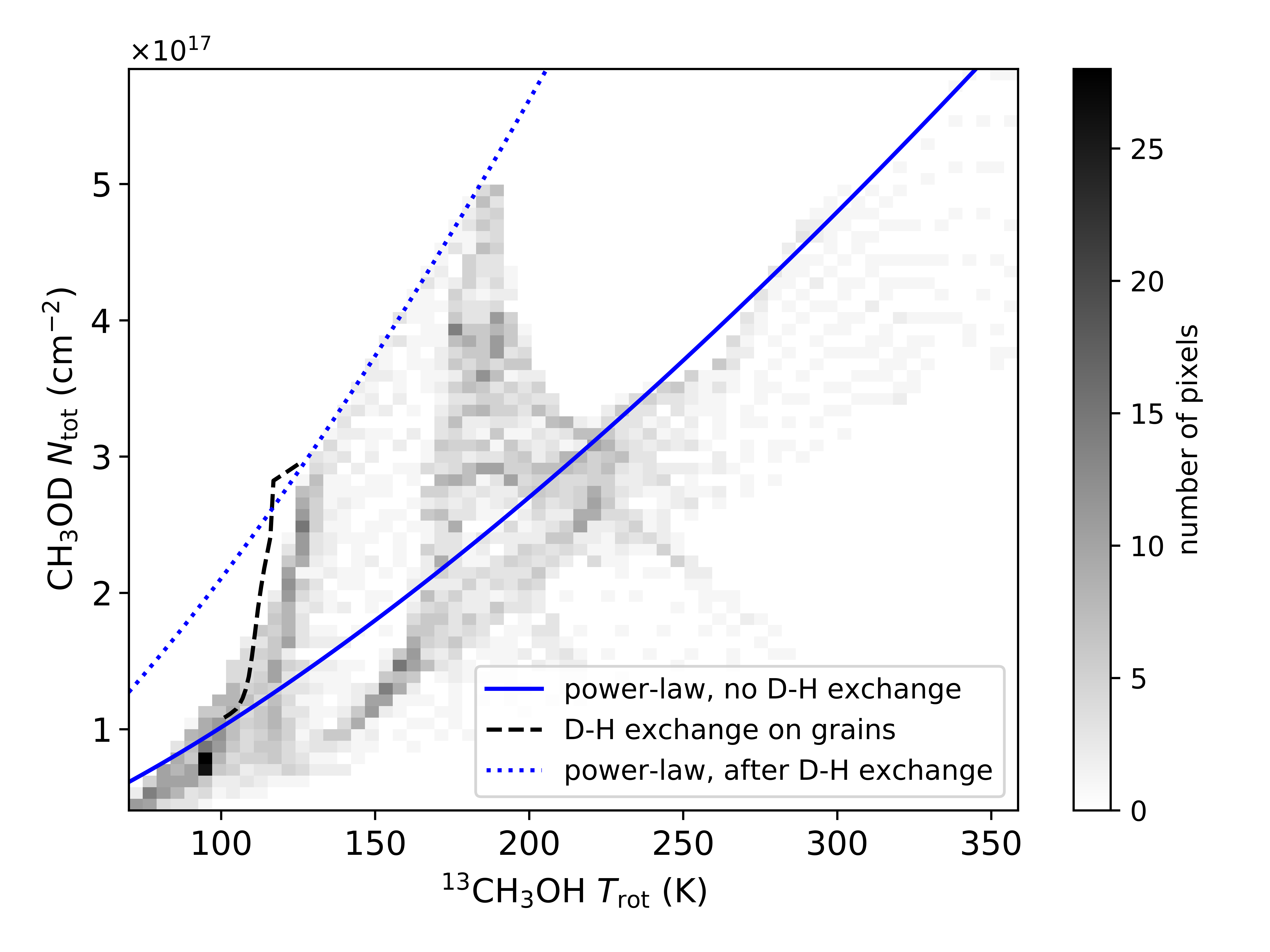}
    \caption{}
    \end{subfigure}    
    \begin{subfigure}{\textwidth}\centering
    \includegraphics[width=0.6\textwidth]{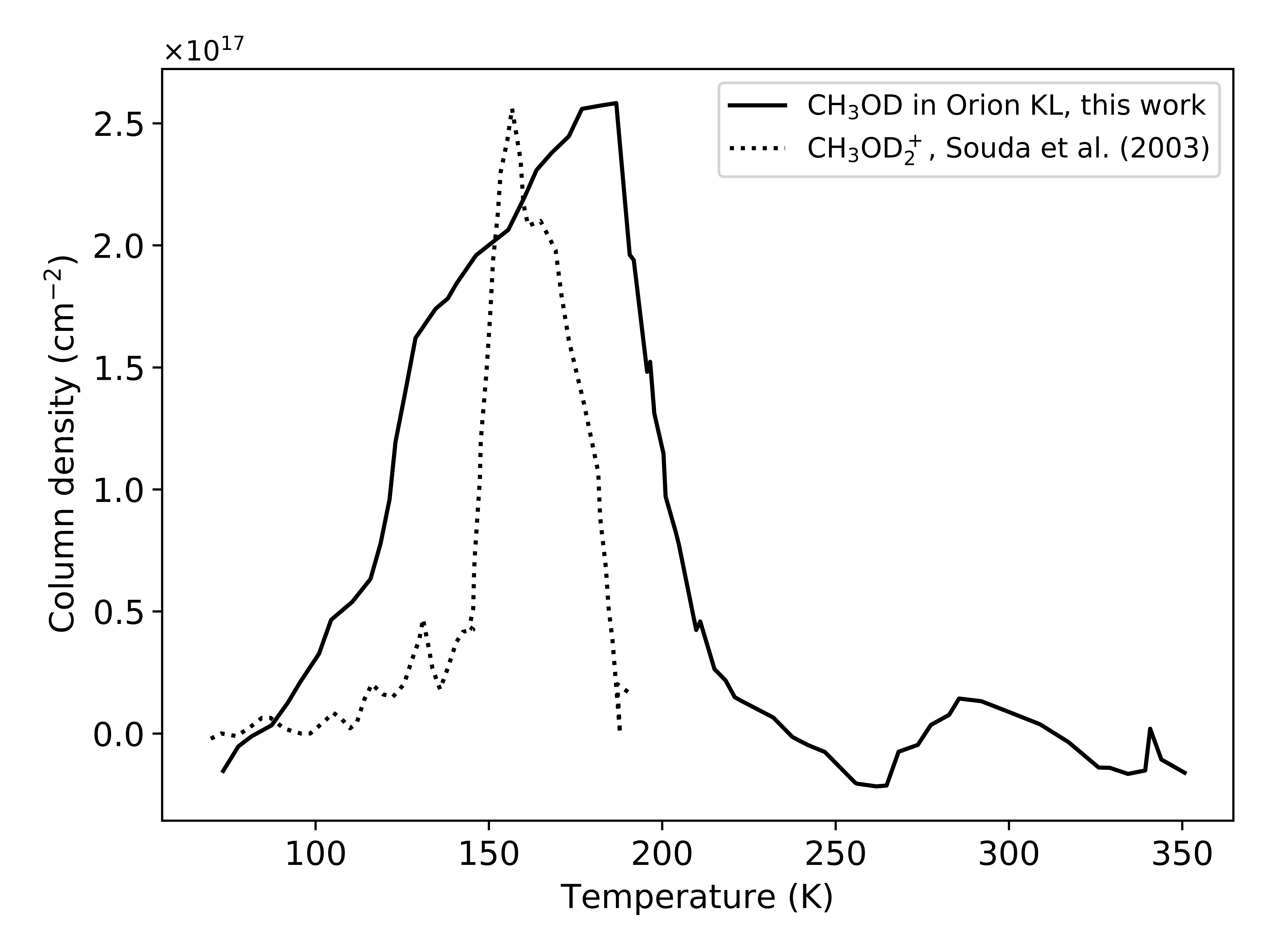}
    \caption{}
    \end{subfigure}
    \caption{(a) Two-dimensional histogram (50 points per bin) showing the derived \ch{CH3OD} column densities $N_{\rm tot}$ 
    against rotational temperature $T_{\rm rot}$ \textcolor{black}{of \ch{^{13}CH3OH}}. The blue solid line shows the power-law fit to the data if there were no \textcolor{black}{D-H exchange} \textcolor{black}{(eq~\ref{eq:powerlaw})}. The black dashed line shows the modeled \textcolor{black}{D-H exchange} \textcolor{black}{(eq~\ref{eqn:soudafit})} followed by desorption based on ice experiments by \citet{Souda2003} and assumptions by \citet{Faure2015dhexchange} The blue dotted curve is the power-law fit after the \ch{CH3OD} enhanced by \textcolor{black}{D-H exchange thermally} desorbs off the grains. 
    (b) The solid curve shows the density profile with the underlying power-law (blue curve in a\textcolor{black}{, eq~\ref{eq:powerlaw}}) subtracted. The dotted curve shows the sputtered \ch{CH3OD2+} \textcolor{black}{profile} reported by \citet{Souda2003} \textcolor{black}{but scaled for comparison to the \ch{CH3OD} column density profile in this work.}}
    \label{fig:nvt}
\end{figure}

Figure~\ref{fig:nvt}a\textcolor{black}{, in which each point represents 50 binned pixels for which a column density and rotational temperature pair were derived,} shows that the \ch{CH3OD} column density increases with rotational temperature, which is characteristic of thermal desorption in which material \textcolor{black}{is sublimed} from the grains as the environment warms. \citep[][]{Burke2010} However, the profile also contains a ``shark-tooth'' feature where the column density starts to rise more \textcolor{black}{steeply} at $\sim$110 K before peaking close to 185 K. At temperatures higher than 185 K, there is a sharp decrease in the \ch{CH3OD} column density, \textcolor{black}{and the relationship between column density and rotational temperature returns to the underlying trend}.


Power-law distributions are commonly used to characterize the temperature and density profiles of star-forming regions and young stellar objects\textcolor{black}{ (YSOs)}.\citep{Carney2017,Gieser2021} 
In Figure~\ref{fig:nvt}a, the fitted underlying power-law relationship between $T_{\rm rot}$ and $N_{\rm tot}$, described by eq \ref{eq:powerlaw}, is shown by the solid blue line.
\begin{equation}\label{eq:powerlaw}
N_{\rm tot}=1.5\times10^{14} T_{\rm rot}^{1.4148}    
\end{equation}
The solid black line in Figure~\ref{fig:nvt}b shows the ``shark-tooth'' from Figure~\ref{fig:nvt}a with the underlying power-law between $T_{\rm rot}$ and $N_{\rm tot}$ subtracted.

\subsection{Grain-Surface Processes}

The rapid rise in gas-phase \ch{CH3OD} column density between $\sim$110 K and $\sim$120 K is consistent with \textcolor{black}{D-H exchange} between methanol and heavy water (HDO, \ch{D2O}) on the ices at $\sim$100 K. \citet{Souda2003} experimentally investigated hydrogen bonding between water and methanol in low-temperature ices warmed from 15 K to 200 K under ultrahigh vacuum conditions. They observed that when \ch{CH3OH} was adsorbed onto \ch{D2O} ice, secondary \ch{CH3OD2+} ions---evidence of \textcolor{black}{D-H exchange} at the hydroxyl site---sputtered off the ice analogue surfaces predominantly between 140 and 175 K. Follow-up analyses by \citet{Kawanowa2004} describes this as a ``rapid and almost complete H/D exchange'' to yield the sputtered \ch{CH3OD2+} species. The fact that we see a similar sudden increase in \ch{CH3OD} column densities at similar temperatures (Figure~\ref{fig:nvt}b, dotted line), with discrepancies owing to the differences in pressure between ultrahigh vacuum and even the densest regions of interstellar medium, supports a similar rapid exchange in Orion KL. 

Models of \textcolor{black}{D-H exchange} between water and methanol in ices by \citet{Faure2015dhexchange} successfully \textcolor{black}{reproduced} gas-phase \ch{CH3OH} deuterium fractionation in Orion KL using initial ice abundances of $n_S(\ch{CH3OH}) = 2.0\times 10^{-6}n_{\ch{H}}$, $n_S(\ch{HDO}) = 3.0\times 10^{-7}n_{\ch{H}}$, and $n_S(\ch{CH3OD}) = 6.0\times 10^{-9}n_{\ch{H}}$. \textcolor{black}{Taking these initial ice abundances, we modeled the change in gas-phase \ch{CH3OD} column density following rapid D-H exchange on the ices and subsequent desorption. Specifically, we assumed an initial ice column density of $N_S(\ch{CH3OD})= 6.0\times10^{-9}N_{\ch{H}} = 6.0\times10^{14} \mbox{ cm}^{-2}$, since $N_{\ch{H}}\sim10^{23}$ cm$^{-2}$ \textcolor{black}{across} Orion KL \textcolor{black}{(including the Hot Core, Compact Ridge, and Extended Ridge)},\citep[][]{Neill2013water,Crockett2014} and an initial gas-phase column density of $N(\ch{CH3OD})=1\times10^{16} \mbox{ cm}^{-2}$, based on the column densities measured at 100 K in this work \textcolor{black}{after subtracting the underlying power-law in eq~\ref{eq:powerlaw}}.}

The enhancement of \textcolor{black}{gas-phase} \ch{CH3OD} column density from \textcolor{black}{D-H exchange} with water was then modeled. 
In the absence of directly analogous temperature-programmed desorption measurements of \ch{CH3OH} and \ch{CH3OD} themselves, we fit the \ch{CH3OD2+} curve of \citet{Souda2003} (Figure~\ref{fig:nvt}b, dotted line) between 110 K and 145 K. The relative intensity amplitude was normalized to the column densities observed for \ch{CH3OD} in our Orion KL observations, resulting in a relationship of the form
\begin{equation}\label{eqn:soudafit}
    N_S^{\prime}(\ch{CH3OD})=6.2\times 10^{14} T -6.0\times10^{16}
\end{equation}
where $N_S^{\prime}$ is the additional (solid) \ch{CH3OD} available [cm$^{-2}$] for desorption at a given temperature $T$. 
The desorption rate coefficient is expressed as 
\begin{equation}\label{eqn:desorptionrate}
k_{des}=\nu_{des}e^{(-E_d/T)}
\end{equation}
where the pre-exponential factor $\nu_{des}$ and the binding energy $E_d$ are taken to be approximately the values for annealed amorphous solid water---$2.0\times10^{12}$ s$^{-1}$ and 5200 K, respectively---under the assumption that the methanol desorbs with water, which is in excess.\citep{Sandford1988,Fraser2001,Brown2007,Dulieu2013,Faure2015dhexchange,Penteado2017}

\textcolor{black}{It follows that the change in gas-phase $N(\ch{CH3OD})$ is approximated by multiplying the rate coefficient (eq~\ref{eqn:desorptionrate}) by the total solid \ch{CH3OD} column density, which includes the additional solid \ch{CH3OD} available following D-H exchange (eq~\ref{eqn:soudafit}) at temperature $T$. Thus, the rate at which $N(\ch{CH3OD})$ changes in the gas phase between 100 and 150 K is approximated by}
\begin{equation}\label{eqn:desrate}
    \frac{d}{dt}N(\ch{CH3OD}) = k_{des}\left[N_S(\ch{CH3OD})+N_S^\prime(\ch{CH3OD})\right]
\end{equation}
\textcolor{black}{using temperature steps of 1 K, the corresponding} time steps for which were determined by 
\begin{equation}\label{eqn:temptime}
    \Delta t = \left(\frac{T-T_0}{T_{max}-T_0}\right)^{1/n}t_h
\end{equation}
where $\Delta t$ is the time elapsed since $t=0$; $T_0$ and $T_{max}$ are the initial (10 K) and maximum (\textcolor{black}{300} K) temperatures, respectively; $t_h$ is the heating timescale; and $n$ is the order of heating, which is assumed to be 2 following the previous work.\citep{Taniguchi2019} \textcolor{black}{The initial gas and solid \ch{CH3OD} column densities were assumed, respectively, to be $N(\ch{CH3OD})=9.0\times 10^{15}$ cm$^{-2}$ (approximated from the \ch{CH3OD} density profile with the underlying power-law subtracted) and $N_S(\ch{CH3OD})=6.0\times10^{-9} N_{\ch{H}}$ cm$^{-2}$ (based on assumptions used by \citet{Faure2015dhexchange} in their D-H exchange models).}

The resulting desorption model from eq~\ref{eqn:desrate} with $t_h=10^3$ yr is shown by the black dashed line in Figure~\ref{fig:nvt}a. Longer timescales (i.e., $t_h\ge10^4$ yr) characteristic of massive YSOs do not follow the increasing \ch{CH3OD} profile as closely. Although potential internal heating sources have been suggested within the Hot Core-SW,\citep[][]{Wilkins2022} \citet{Li2020} conclude that this region, part of the ``elongated ridge'' comprising the Hot Core and Source I, is predominantly heated externally by shocks induced by the Orion KL explosion, which took place about 500 years ago. Furthermore, the models by \citet{Faure2015dhexchange} suggest the \textcolor{black}{D-H exchange} in Orion KL reaches steady-state in ${<}10^3$ yr. 
After \textcolor{black}{D-H exchange}, the power-law relationship from eq~\ref{eq:powerlaw} applies, but with a coefficient of $3.2\times 10^{14}$ \textcolor{black}{(dotted blue curve in Fig.~\ref{fig:nvt}a)}.

The hypothesis that the \ch{CH3OD} column density profile in Figure~\ref{fig:nvt} is the result of rapid \textcolor{black}{D-H exchange} on the grains relies on two assumptions regarding temperature. First, we assume that the rotational temperature $T_{\rm rot}$ is an appropriate estimate for the kinetic temperature $T_{kin}$. This assumption is based on the fact that the Compact Ridge\textcolor{black}{, a spatial component toward the southwestern region of Orion KL that is characterized as being rich in oxygen-bearing molecules,} has a fairly high density of ${\sim}10^6$ cm$^{-3}$, \citep{Blake1987,Genzel1989} implying that LTE is a reasonable assumption. Second, we assume that the dust and gas are thermally coupled. \citet{Li2003} found that other quiescent regions (no infrared sources, no evident outflows) in the Orion Molecular Cloud are thermally coupled. Models by \textcolor{black}{\citet{Bruderer2009}} also support coupling between dust and gas temperature at \textcolor{black}{the relevant} densities. \textcolor{black}{This gas-grain thermal coupling was also demonstrated in models of several massive star-forming regions, including the Orion KL Compact Ridge, by \citeauthor{Garrod2006coupling}.\citep[][]{Garrod2006coupling}} As such, it is reasonable to assume that the gas temperatures shown in Figure~\ref{fig:nvt} are also representative of the temperatures of the dust, on which methanol forms, and that any decoupling between the dust and gas is negligible for the purposes of this discussion. 

\subsection{Enhanced Deuteration in the Hot Core-SW}

\begin{figure}
    \centering
    \begin{subfigure}{0.495\textwidth}
    \includegraphics[width=\textwidth]{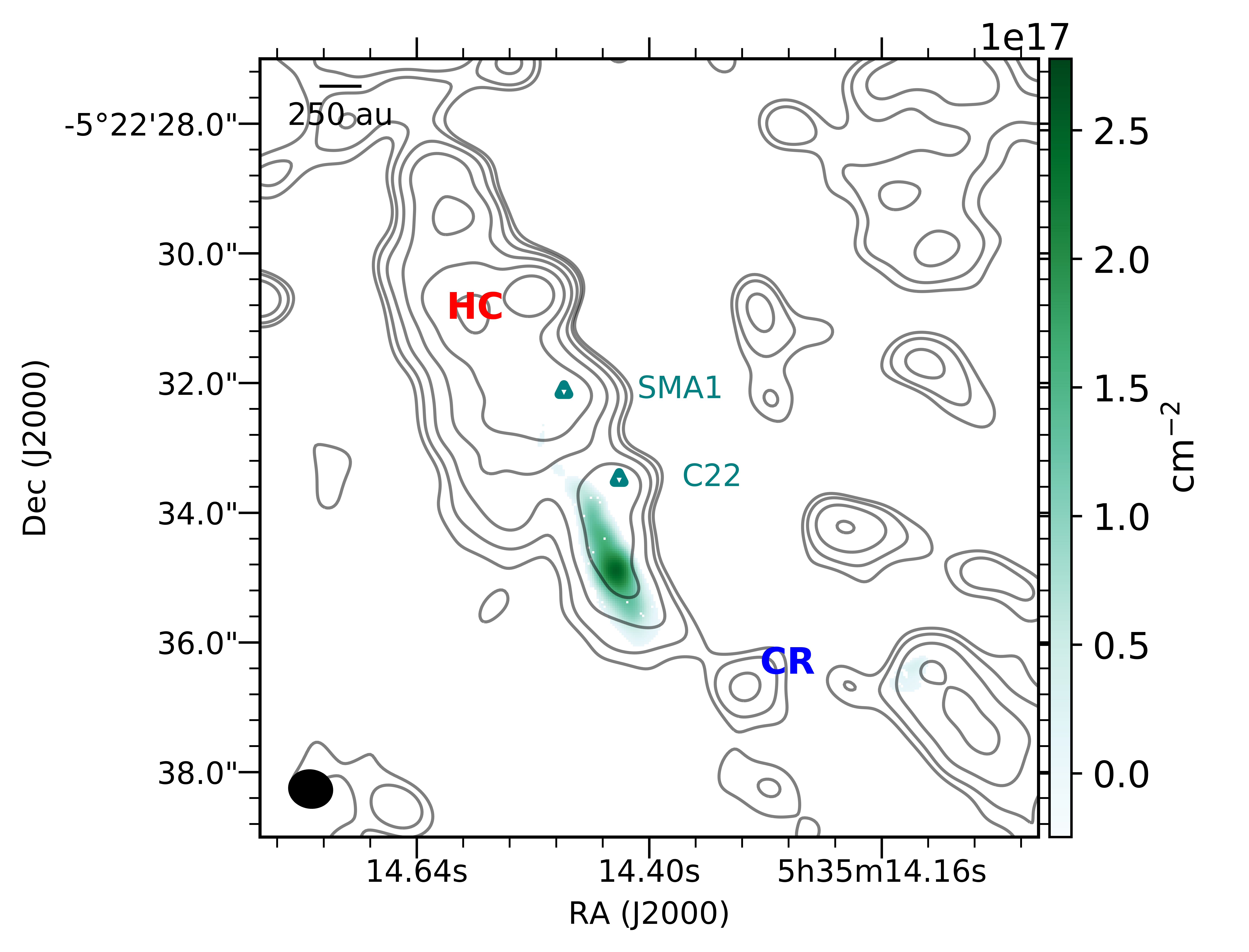}
    \caption{}
    \end{subfigure}
    \begin{subfigure}{0.495\textwidth}
    \includegraphics[width=\textwidth]{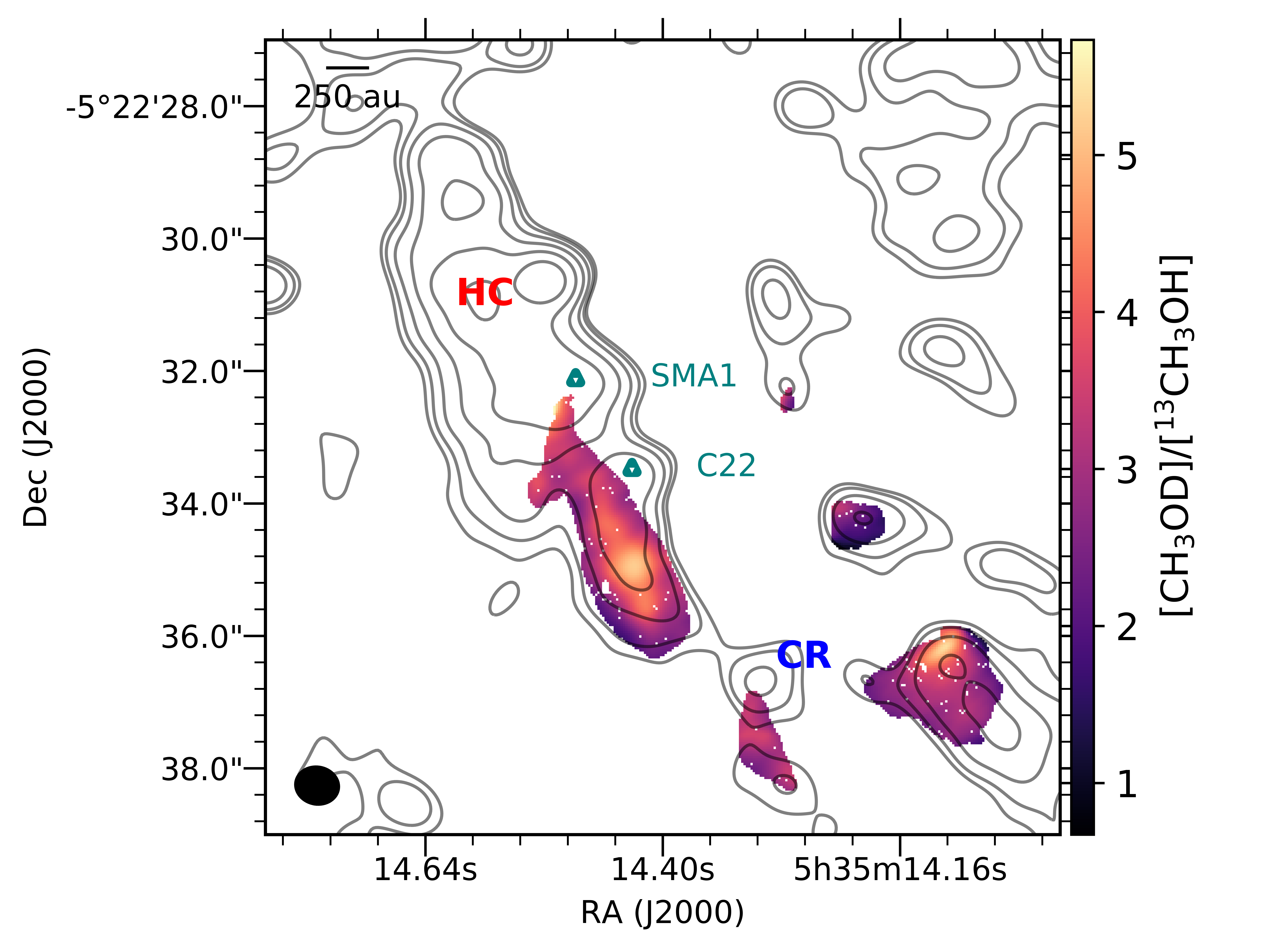}
    \caption{}
    \end{subfigure}
    \caption{\textcolor{black}{(a)}Map of \textcolor{black}{gas-phase }\ch{CH3OD} \textcolor{black}{excess} column density across Orion KL after subtracting the underlying power-law relationship (eq~\ref{eq:powerlaw}) between rotational temperature and column density. Darker shading indicates a larger deviation from the underlying power-law. \textcolor{black}{(b) Map of [\ch{CH3OD}]/[\ch{^{13}CH3OH}].}}
    \label{fig:sharkmap}
\end{figure}

Above $\sim$180 K, the profile in Figure~\ref{fig:nvt}a drops sharply. \textcolor{black}{We attribute this to the enhanced deuteration being localized to the Hot Core-SW and that there simply is little gas above 180 K in this region.} Figure~\ref{fig:sharkmap}a shows the distribution of \ch{CH3OD} column densities minus the underlying power-law relationship (eq~\ref{eq:powerlaw}). Figure~\ref{fig:sharkmap}b maps [\ch{CH3OD}]/[\ch{^{13}CH3OH}], confirming that the profile in Figure~\ref{fig:sharkmap}a is indeed the result of excess \ch{CH3OD} in the Hot Core-SW and not enhanced abundances of methanol in general.

\textcolor{black}{We considered two chemical explanations for this trend, neither of which adequately explain the observed patterns.} One avenue \textcolor{black}{that has been proposed for} \ch{CH3OD} depletion is gas-phase \textcolor{black}{D-H exchange} via protonation of the hydroxyl group (eqs~\ref{eq:protonationforreal} \textcolor{black}{and~\ref{eq:protonationforreal2}}) followed by dissociative recombination (eqs.~\ref{eq:protonation} and~\ref{eq:protonation2}).\citep[][]{Osamura2004}.
\begin{equation}\label{eq:protonationforreal}
\ch{CH3OD + H3O+ -> CH3ODH+ + H2O}
\end{equation}
\begin{equation}\label{eq:protonationforreal2}
\textcolor{black}{\ch{CH2DOH + H3O+ -> CH2DOH2+ + H2O}}
\end{equation}
\begin{equation}\label{eq:protonation}
\ch{CH3ODH+ + e- -> CH3OH + D}, \ch{CH3OD + H}
\end{equation}
\begin{equation}\label{eq:protonation2}
\ch{CH2DOH2+ + e- -> CH2DOH + H}
\end{equation}
The methyl H/D site is not exchangeable; therefore, in this model, only \ch{CH3OD} can be depleted while \ch{CH2DOH} cannot, which has been suggested as an explanation for the low relative \ch{CH3OD} abundances in low-mass star-forming regions.\citep[][]{Osamura2004} \textcolor{black}{However, this is unlikely to account for the decrease in gas-phase \ch{CH3OD} at warmer temperatures in the Orion KL Hot Core-SW following enrichment on grain surfaces because dissociative recombination reactions tend to destroy gas-phase methanol (and its isotopologues), with methanol production comprising the smallest branching ratio (3\%) listed in KIDA\citep[][]{KIDA}. A drop in the \ch{^{13}CH3OH} column density at temperatures above 180 K is not seen in Orion KL (see Figure S5 in the Supporting Information), suggesting that the observed decrease of the [\ch{CH3OD}]/[\ch{^{13}CH3OH}] ratio (Figure~\ref{fig:DvsHbyTemp}b) and of the \ch{CH3OD} column density (Figure~\ref{fig:nvt}a) above 180 K cannot simply be explained by protonation of the methanol (e.g., by \ch{H3O+}) followed by dissociative recombination.}

Another mechanism considered was the neutral-radical reaction with the hydroxyl radical (OH); however, this reaction is too slow at $\sim$200 K to account for the observed patterns (see Section S4 of the Supporting Information).

\textcolor{black}{Instead, we propose that the drop in enhanced \ch{CH3OD} column density is the result of environment rather than chemistry. That is, the enhanced \ch{CH3OD} column density profile is limited to warm gas in the Hot Core-SW with temperatures that max out around 200 K. The enhanced deuteration in this region, brought about by temperature-dependent surface D-H exchange, may be the result of evolutionary state. However, this region is not directly associated with any known YSOs, including SMA1 (a young, high-mass protostar) and C22 (a possible hot core).\citep[][]{Beuther2006,Friedel2011} Although there is no known self-illuminated source (e.g., embedded protostar) in the Hot Core-SW to drive grain warming and associated D-H exchange in ices, it has been suggested that there is a potential (hidden) source of internal heating there.\citep[][]{Wilkins2022} \textcolor{black}{And, the complex history and star formation patterns in Orion KL will lead to quite varied thermal histories of various sub-regions of the giant molecular cloud complex.} Thus, dedicated work to elucidate the nature of the Hot Core-SW region is needed to better understand how this environment could affect the observed chemistry.
} 

\subsection{Methyl Group Chemistry}\label{sec:methyl}

\begin{figure}
    \centering
    \includegraphics[width=0.7\textwidth]{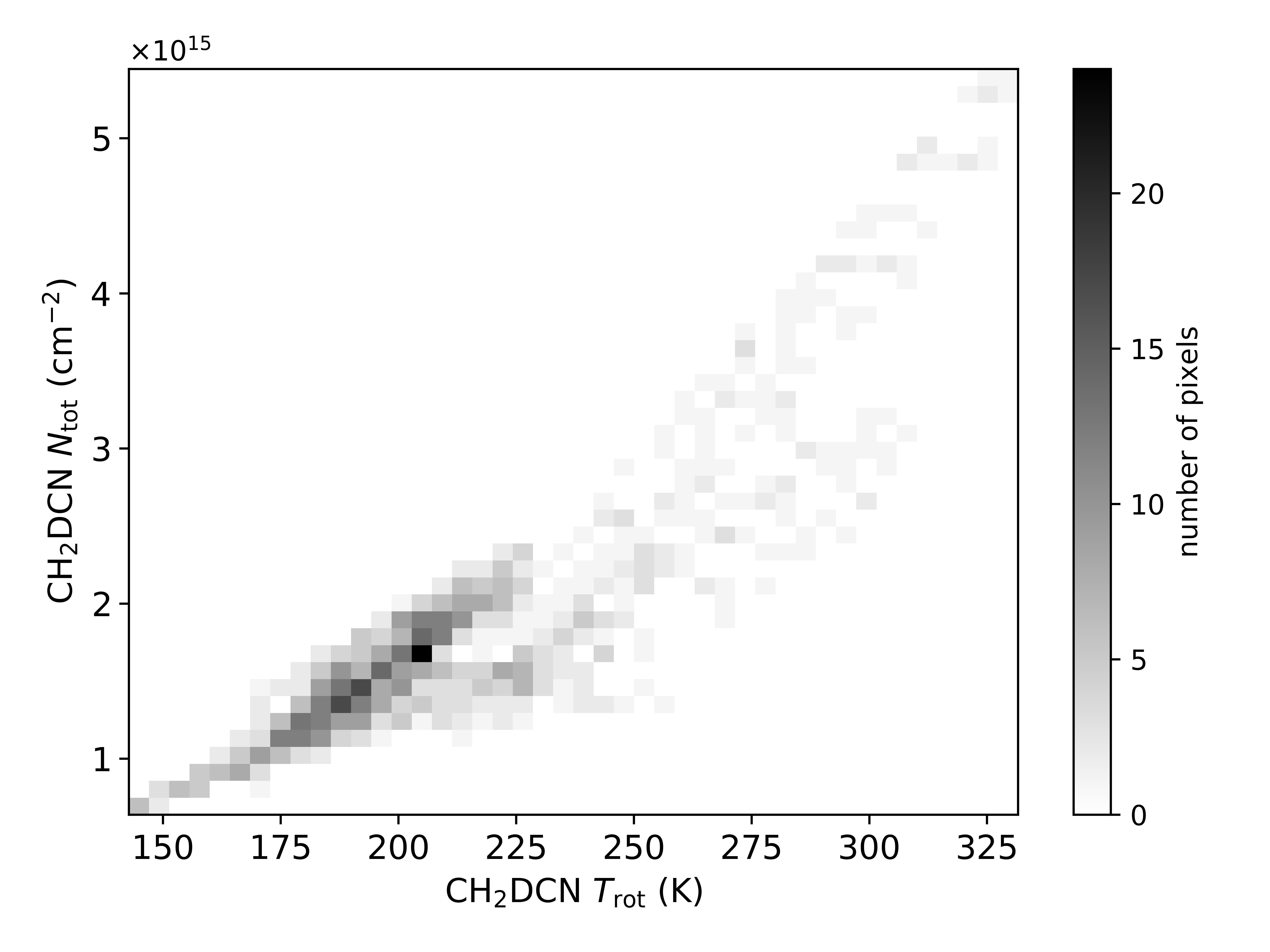}
    \caption[Two-dimensional histogram of \ch{CH2DCN} column density and temperature]{Two-dimensional histogram with 20 points per bin showing the derived \ch{CH2DCN} column density against rotational temperature, using data presented by \citet{Carroll2017thesis}.}
    \label{fig:nvtch2dcn}
\end{figure}

If the hump observed in the \ch{CH3OD} column density versus temperature profile is indeed evidence of surface \textcolor{black}{D-H exchange} at the hydroxyl site, then we would expect a smooth profile (i.e., without a similar hump) in the profile of \ch{CH2DOH}. Unfortunately, we do not have sufficient \ch{CH2DOH} transitions in our data to test this hypothesis directly. However, \citet{Carroll2017thesis} mapped the physical parameters of \ch{CH2DCN} toward Orion KL using data from ALMA (project: ADS/JAO.ALMA\#2013.1.01034\textcolor{black}{, PI: Crockett}). Figure~\ref{fig:nvtch2dcn} shows a histogram of the \ch{CH2DCN} column density and rotational temperature using these data where they overlap with \ch{CH3OD} emission in the current observations. In this plot, we see a consistent power-law relationship between temperature and column density, which supports the conclusion by \citet{Osamura2004} that the methyl groups of complex organics would not undergo grain\textcolor{black}{-surface} or gas-phase \textcolor{black}{D-H exchange}. In other words, the hump visible in Figure~\ref{fig:nvt}a is indeed likely the result of chemistry specific to the hydroxyl site of methanol. \textcolor{black}{Future dedicated observations of \ch{CH2DOH} lines with ALMA in this region are necessary to confirm this hypothesis.} 

\subsection{Comparison to Past Studies}\label{sec:comparison}

The evidence presented here adds to the growing list of observational and theoretical evidence in favor of a grain-surface mechanism for \ch{CH3OD} enrichment in massive star-forming regions. A key difference between this work and that of past observations is that, here, we map \ch{CH3OD} column density across much of Orion KL, including the Compact Ridge, whereas past work derives one value for the Compact Ridge as a whole. Furthermore, resolving the small-scale structure of \ch{CH3OD} column densities (and temperature) allows us to look at how column density is related to the line-of-site temperature, something that has not yet been extensively investigated through observations but is now possible with the sensitivity and \textcolor{black}{spatial} grasp of ALMA.

Computational models to assess \textcolor{black}{D-H exchange} generally have investigated {\em temporal} variations in relative \ch{CH3OD} abundances at a single temperature or have looked at singly-deuterated methanol chemistry across a range of temperatures but less than 140 K.\citep{Osamura2004,Faure2015dhexchange,Bogelund2018} \textcolor{black}{Previous models of grain-surface D-H exchange in Orion KL have assumed methanol is completely sublimated at temperatures $>$110 K, whereas our observations suggest D-H exchange on the ice may be important up to 125 K.} Thus, the observations presented here probe a temperature regime beyond that of existing astrochemical models and call for revised models to investigate D/H chemistry at higher temperatures. \textcolor{black}{Specifically, temperature-dependent models of 
grain-surface D-H exchange in which methanol sublimates completely at higher temperatures (e.g., 125 K) 
are needed to more robustly explain the patterns observed in this work.}

Perhaps the leading criticism of proposed grain-surface chemistry prompting the enhancement of \ch{CH3OD} abundances relative to \ch{CH2DOH} is that such processes would require a large initial [HDO]/[\ch{H2O}] ratio\textcolor{black}{. For example, models by \citet{Charnley1997} suggest that the initial [HDO]/[\ch{H2O}] ratio in the ice mantles would need to be} $\sim$0.1,\citep{Charnley1997} which is significantly larger than the ratio of $\sim$0.003 reported by \citet{Neill2013water} for compact regions of Orion KL. However, \citet{Thi2010} suggest that the [HDO]/[\ch{H2O}] ratio can exceed 0.01 in dense (${\ge}10^6$ cm$^{-3}$), warm ($T>100$ K) regions (such as those observed here) via neutral-neutral reactions\textcolor{black}{---such as the formation of \ch{HDO} from \ch{OH + HD}, \ch{OD + H2}, and \ch{OD + OH}---}which may be promising for the hypothesis of a grain-surface \ch{CH3OD} enhancement.  Even more promising is a model presented by \citet{Faure2015dhexchange}, who reproduced observed [\ch{CH2DOH}]/[\ch{CH3OD}] ratios in the Compact Ridge assuming a primitive [HDO]/[\ch{H2O}] fractionation of 0.006, only a factor of 2 larger than the observed ratio reported by \citet{Neill2013water}. That is, the key initial condition is that of the deuteration state of methanol and water in the icy grain mantles, which are exceedingly difficult to measure directly with previous observational capabilities. JWST will offer greatly improved capabilities to attempt such measurements, going forward.

A remaining question, then, is what makes methanol deuteration in low-mass star-forming regions so different from that in high-mass star-forming regions? 
\citet{Ratajczak2011} suggest observational biases, namely that, since high-mass objects tend to be further away than those low-mass objects where deuterium chemistry has been studied, measurements of the [\ch{CH2DOH}]/[\ch{CH3OD}] ratio may be affected, particularly if the spatial distributions of the two isotopomers are different. High angular resolution mapping \textcolor{black}{of high-mass star-forming regions}, like that presented in Figure~\ref{fig:columnD}, would address this by comparing the \ch{CH2DOH} and \ch{CH3OD} column densities only where their emission overlapped\textcolor{black}{, as was done for different spectral components of the high-mass star-forming region NGC 6334I by \citet{Bogelund2018}} As stated previously, 
the observations presented here do not have sufficient \ch{CH2DOH} lines available to test \textcolor{black}{the spatial correlation of site-specific deuterated isotopologues, and} dedicated high angular resolution observations targeting low-energy \ch{CH2DOH} lines are necessary to further address \textcolor{black}{this issue}.

Another conjecture for the different deuterium fractionation patterns in massive YSOs compared to low-mass star-forming regions is that there is simply less deuteration in massive protostars because of the warmer environments.\citep[][]{Bogelund2018} \citet{Faure2015dhexchange} reproduced relative singly-deuterated methanol abundances for Orion KL (a high-mass source) and IRAS 16293-2422 (a low-mass object) using kinetic models that were identical except for the initial deuterium fractionation ratios. They reported that the Orion KL Compact Ridge's gas-phase deuterium chemistry could be modeled assuming similar primitive deuteration of water and methanol ices ($\sim$0.2-0.3\%), whereas IRAS 16293's gas-phase deuterated methanol chemistry required a significantly higher deuterium fractionation in methanol (12\%) than water (1\%). 
Their model shows complete methanol desorption by $\sim$110 K. Such conditions of extreme deuteration only occur in very cold, dense environments where extensive molecular depletion occurs, including that of CO and N$_2$. Under such conditions, D$_3 ^+$ becomes the dominant molecular ion, whose dissociative recombination results in the arrival of \textcolor{black}{hydrogen }atoms onto grain mantles \textcolor{black}{with a D/H} ratio of $>$1.

As seen in Figure~\ref{fig:nvt}a, the desorption model based on work by \citet{Faure2015dhexchange} (dashed line) matches nicely the \ch{CH3OD} column density rise between 100 and 110 K; however, the \ch{CH3OD} in our data increases at temperatures up to $\sim$125 K. This slight discrepancy might be addressed by temperature-programmed desorption experiments, for example, studying the release of \ch{CH3OD} directly rather than via sputtered \ch{CH3OD2+} detected by \citet{Souda2003} \textcolor{black}{Furthermore, thermal  desorption strongly depends on the composition of the underlying surface.} Such questions require more robust chemical networks for deuterium chemistry as well as a better understanding of the initial chemical conditions of both high-mass and low-mass star-forming regions.

\section{Conclusion}\label{sec:summarych3od}

We provide observational evidence in support of rapid \textcolor{black}{D-H exchange} in methanol\textcolor{black}{-containing ices}, specifically at the hydroxyl site, between $\sim$100 and 125 K in Orion KL, using high angular resolution ALMA Band 4 observations of \ch{CH3OD} to map the small-scale variations in \ch{CH3OD} column density for the first time, and to compare the observed column densities to the line-of-site rotational temperatures mapped at the same angular resolution (and derived previously from \ch{^{13}CH3OH}).\citep{Wilkins2022} 

We fit power-law relationships and toy models of \textcolor{black}{D-H exchange} at methanol's hydroxyl (-OH) site followed by \ch{CH3OD} \textcolor{black}{thermal} desorption to the observed \ch{CH3OD} column density profile in Orion KL. These analyses suggest that \textcolor{black}{D-H exchange} and rapid \ch{CH3OD} desorption increase the gas-phase \ch{CH3OD} column density between 100 and 125 K. \textcolor{black}{Enhanced \ch{CH3OD} column densities are limited to the Hot Core-SW, which has been previously suggested to harbor a potential source of internal heating, which could explain the enhanced \ch{CH3OD} column densities between 125 and 185 K. In this interpretation, there is simply little gas in this region at higher temperatures that would display \ch{CH3OD} enhancements.} 

Future investigations---through observations, experiments, and computational models---are needed to further constrain the peculiar D-methanol chemistry in Orion KL and other star-forming regions. The work presented here would be aided by dedicated high-resolution observations of \ch{CH2DOH}, 
spectroscopic experiments measuring the kinetics of \ch{CH3OD} formation via \textcolor{black}{D-H exchange} in heavy water ice \textcolor{black}{(HDO and \ch{D2O})} and subsequent desorption, and temperature-dependent astrochemical models of possible \ch{CH3OD} loss at higher temperatures (185-225 K). 

\begin{suppinfo}

\textcolor{black}{The Supporting Information includes an explanation of optical depth approximations (Section S1); supplemental maps (Section S2), specifically \ch{CH3OD} integrated intensity (Figure S2), \ch{^{13}CH3OH} column density (Figure S3), and \ch{^{13}CH3OH} rotational temperature (Figure S4) maps; a comparison of derived [\ch{CH3OD}]/[\ch{^{12}CH3OH}] ratios (Section S3); and  loss calculations of \ch{CH3OD} by gas-phase reactions with OH (Section S4). }

\end{suppinfo}

\section*{Acknowledgements}

This work makes use of the following ALMA data: ADS/JAO.ALMA\#2017.1.01149 and ADS/JAO.ALMA\#2013.1.01034.
ALMA is a partnership of ESO (representing its member states), NSF (USA), and NINS (Japan), together with NRC (Canada), MOST and ASIAA (Tawain), and KASI (Republic of Korea), in cooperation with the Republic of Chile. The Joint ALMA Observatory is operated by ESO, AUI/NRAO, and NAOJ. The National Radio Astronomy Observatory (NRAO) is a facility of the National Science Foundation (NSF) operated under Associated Universities, Inc. (AUI). This research made use of APLpy, an open-source plotting package for Python,\citep{aplpy2019} and the KInetic Database for Astrochemistry (KIDA), an online database of kinetic data.\citep{KIDA}

This work has been supported by the NSF Graduate Research Fellowship Program under grant No. DGE-1144469 and NRAO Student Observing Support under Award No. SOSPA6-014. OHW was additionally supported by an ARCS Los Angeles Founder Chapter scholarship. GAB gratefully acknowledges support from the NSF AAG (AST-1514918) and NASA Astrobiology (NNX15AT33A) and Exoplanet Research (XRP, NNX16AB48G) programs. This work benefited from discussions with Brandon Carroll, Dana Anderson, Aida Behmard, Cam Buzard, Steve Charnley, and Catherine Walsh. \textcolor{black}{Many thanks to the anonymous referees for their thoughtful comments and help in improving the manuscript.} OHW thanks Erica Keller, Sarah Wood, and the NRAO North American ALMA Science Center (NAASC) for their assistance with the data reduction.

\newpage
\bibliography{orionbib}

\begin{tocentry}
\includegraphics[width=\textwidth]{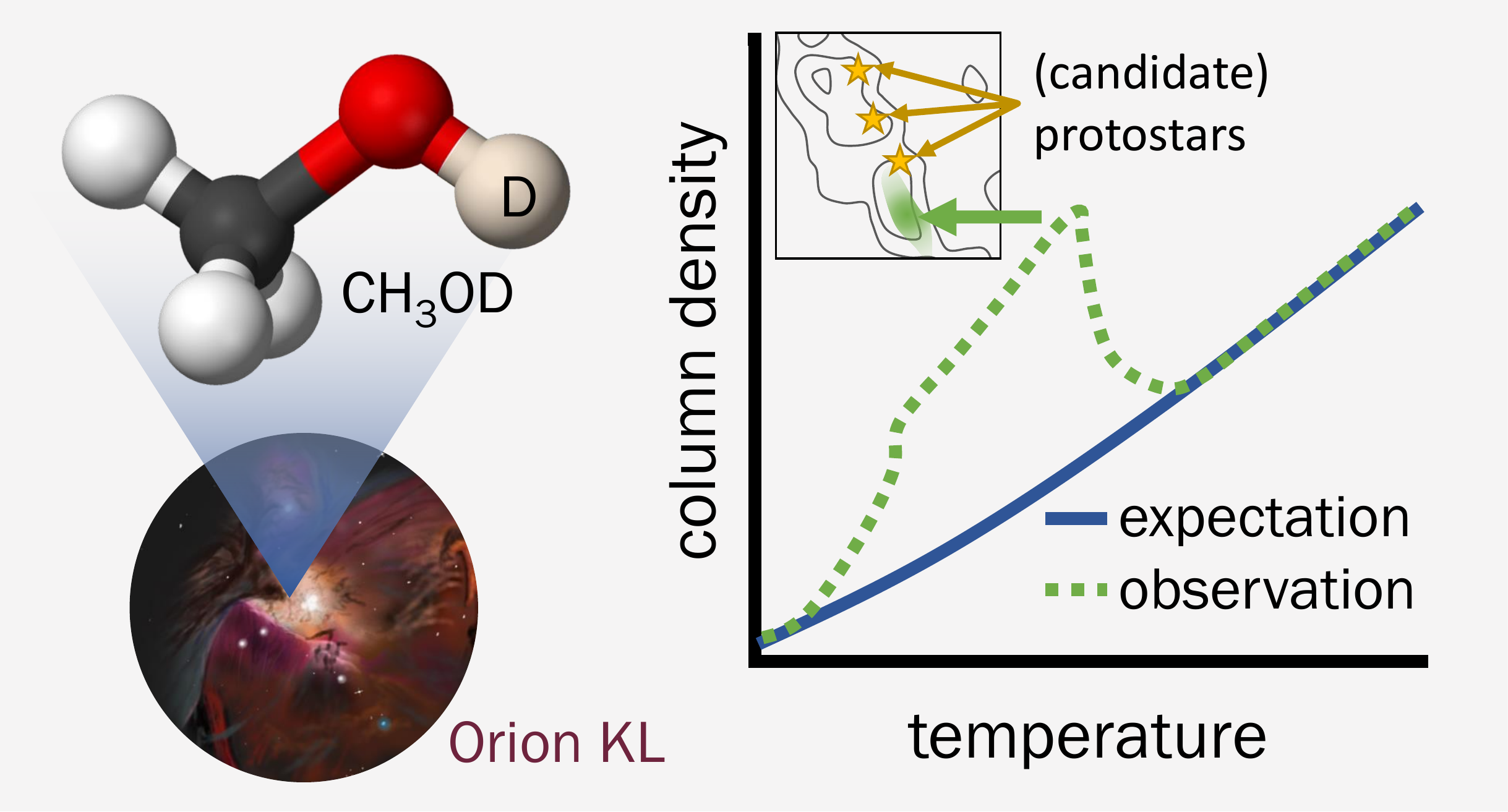}
\end{tocentry}

\end{document}